\documentclass{ws-procs9x6}

\begin{document}

\title{DARK MATTER SEARCHES WITH IMAGING ATMOSPHERIC CHERENKOV TELESCOPES
}

\author{E. MOULIN$^*$}

\address{CEA -Saclay, IRFU,\\
Gif-sur-Yvette, 91191, France\\
$^*$E-mail: emmanuel.moulin@cea.fr\\
}

%

\begin{abstract}
The annihilations of WIMPs produce high energy gamma-rays in the final state.
These high energy gamma-rays may be detected by imaging atmospheric
Cherenkov telescopes (IACTs).
Amongst the plausible targets are the Galactic Center, the centre of galaxy clusters,
dwarf Sphreroidal galaxies and substructures in Galactic haloes.
I will review on the recent results from observations of ongoing IACTs.
\end{abstract}


\bodymatter

\section{Introduction}\label{aba:intro}
Cosmological and astrophysical probes  suggest that $\sim$23\% of the Universe
is composed of non-baryonic dark matter (DM),
commonly assumed to be in the form of Weakly Interacting Massive
Particles (WIMPs) arising in extensions
of the Standard Model of Particle Physics (for reviews see,
e.g.~\cite{Bergstrom:2000pn,Bertone:2004pz}).
Amongst the most widely discussed DM candidates are the lightest neutralino
in supersymmetric extensions of the Standard
Model~\cite{Jungman:1995df} and the first excitation of the Kaluza-Klein
bosons (LKP) in universal extra dimension theories.\cite{appelquist,cheng,servant}

The annihilation of WIMP pairs can produce in the final state a
continuum of gamma-rays whose flux extends up to the DM particle
mass, from the hadronization and decay of the cascading annihilation
products. In supersymmetric models, the gamma-ray spectrum from
neutralino annihilation is not uniquely determined and the branching
ratios (BRs) of the open annihilation channels are not determined
since the DM particle field content is not known {\it a priori}. In
contrast, in Kaluza-Klein scenarios where the lightest Kaluza-Klein
particle (LKP) is the first KK mode of the hypercharge gauge boson,
the BRs of the annihilation channels can be computed given that the
field content of the DM particle is known. The gamma-ray flux from
annihilations of DM particles of mass $m_{DM}$ accumulating in a
spherical DM halo can be expressed in the form :
\begin{equation}
\label{eqnp}
\frac{d\Phi(\Delta\Omega,E_{\gamma})}{dE_{\gamma}}\,=\frac{1}{8\pi}\,\underbrace{\frac{\langle
\sigma
v\rangle}{m^2_{DM}}\,\frac{dN_{\gamma}}{dE_{\gamma}}}_{\rm Particle\,
Physics}\,\times\,\underbrace{\bar{J}(\Delta\Omega)\Delta\Omega}_{\rm Astrophysics}
\end{equation}
as a product of a particle physics component with an astrophysics
component. The particle physics part contains $\langle \sigma
v\rangle$, the velocity-weighted annihilation cross section, and
$dN_{\gamma}/dE_{\gamma}$,  the differential gamma-ray spectrum
summed over the whole final states with their corresponding
branching ratios. The astrophysical part corresponds to the
line-of-sight-integrated squared density of the DM distribution J,
averaged over the instrument solid angle
$\Delta\Omega$ usually matching the angular resolution of the instrument :
\begin{equation}
\label{eqnj} J\,=\,\int_{l.o.s}\rho^2(r[s])ds \hspace{2cm}
\bar{J}(\Delta\Omega)\,=\,\frac{1}{\Delta\Omega}\,\int_{\Delta\Omega}
\,PSF\times J\, d\Omega
\end{equation}
where $PSF$ stands for the point spread function of the instrument.

The annihilation rate being proportional to the square of the DM
density integrated along the line of sight, regions with enhanced DM
density are primary targets for indirect searches. Among them are
the Galactic halo, external galaxies, galaxy clusters, substructures
in galactic haloes, and the Galactic Center. We report here  on
recent results on dark matter searches with current IACTs such as
H.E.S.S. and  MAGIC, towards the Galactic Center, dwarf galaxies
from  the Local Group, Galactic globular clusters and DM mini-spikes
around intermediate mass black holes (IMBHs).

\section{The Galactic Centre}\label{aba:gc}

H.E.S.S. observations towards the Galactic Center have revealed a
bright pointlike gamma-ray source, HESS
J1745-290~\cite{Aharonian:2006wh}, coincident in position with the
supermassive black hole Sgr A*, with a size lower than 15 pc.
Diffuse emission along the Galactic plane has also been
detected~\cite{Aharonian:2006au} and correlates well with the mass
density of molecular clouds from the Central Molecular  Zone, as
traced by CS emission~\cite{tsuboi}. According to recent detailed
studies~\cite{:2007kh}, the source position is located at an angular
distance of 7.3$^{\prime\prime}\pm$8.7$^{\prime\prime}_{\rm
stat.}\pm$8.5$^{\prime\prime}_{\rm syst.}$ from Sgr A*. The pointing
accuracy allows to discard the association of the very high energy
(VHE) emission with the center of the radio emission of the
supernova remnant Sgr A East but the association with the pulsar
wind nebula G359.95-0.04 can not be ruled out. From 2004 data set,
the energy spectrum of the source is well fitted in the energy range
160 GeV - 30 TeV to a power-law spectrum dN/dE $\propto$
E$^{-\Gamma}$ with a spectral index  $\Gamma$ = 2.25$\pm$0.04$_{\rm
stat.}$$\pm$0.1$_{\rm syst.}$. No deviation from a power-law is
observed leading to an upper limit on the energy cut-off of 9 TeV
(95\% C.L.). The VHE emission from HESS J1745-290 does not show any
significant periodicity or variability from 10 minutes to 1
year~\cite{:2007kh}. Besides plausible astrophysical origins~ (see
e.g.~\cite{Aharonian:2004jr} and references therein), an alternative
explanation is the annihilation of DM in the central cusp of our
Galaxy. The spectrum of HESS J1745-290 shows no indication for
gamma-ray lines. The observed gamma-ray flux may also result from
secondaries of DM annihilation. The left hand side of
Fig.~\ref{aba:gc}shows the H.E.S.S. spectrum extending up to masses
of about 10 TeV, which requires large neutralino masses ($>$10 TeV).
They are unnatural in phenomenological MSSM scenarios. The
Kaluza-Klein models provide harder spectra which still significantly
deviate from the measured one. Non minimal version of the MSSM may
yield flatter spectrum with mixed 70\% $b\bar{b}$ and 30\%
$\tau^+\tau^-$ final states. Even this scenario does not fit to the
measured spectrum. The hypothesis that the spectrum measured by
H.E.S.S. originates only from DM particle annihilations is highly
disfavored.
\begin{figure}  
\begin{center}
\mbox{
\psfig{file=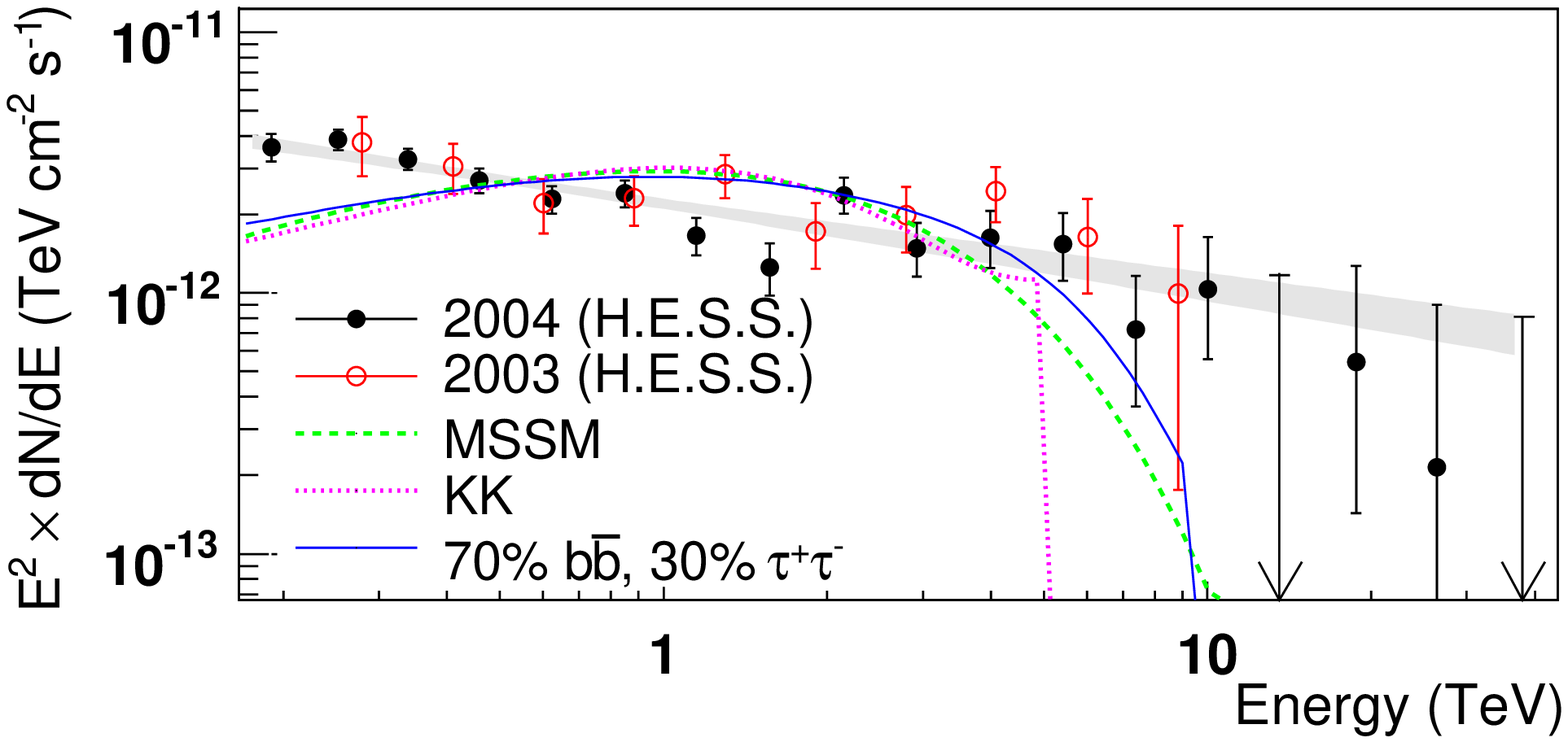,width=5.5cm,height=3.5cm}
\psfig{file=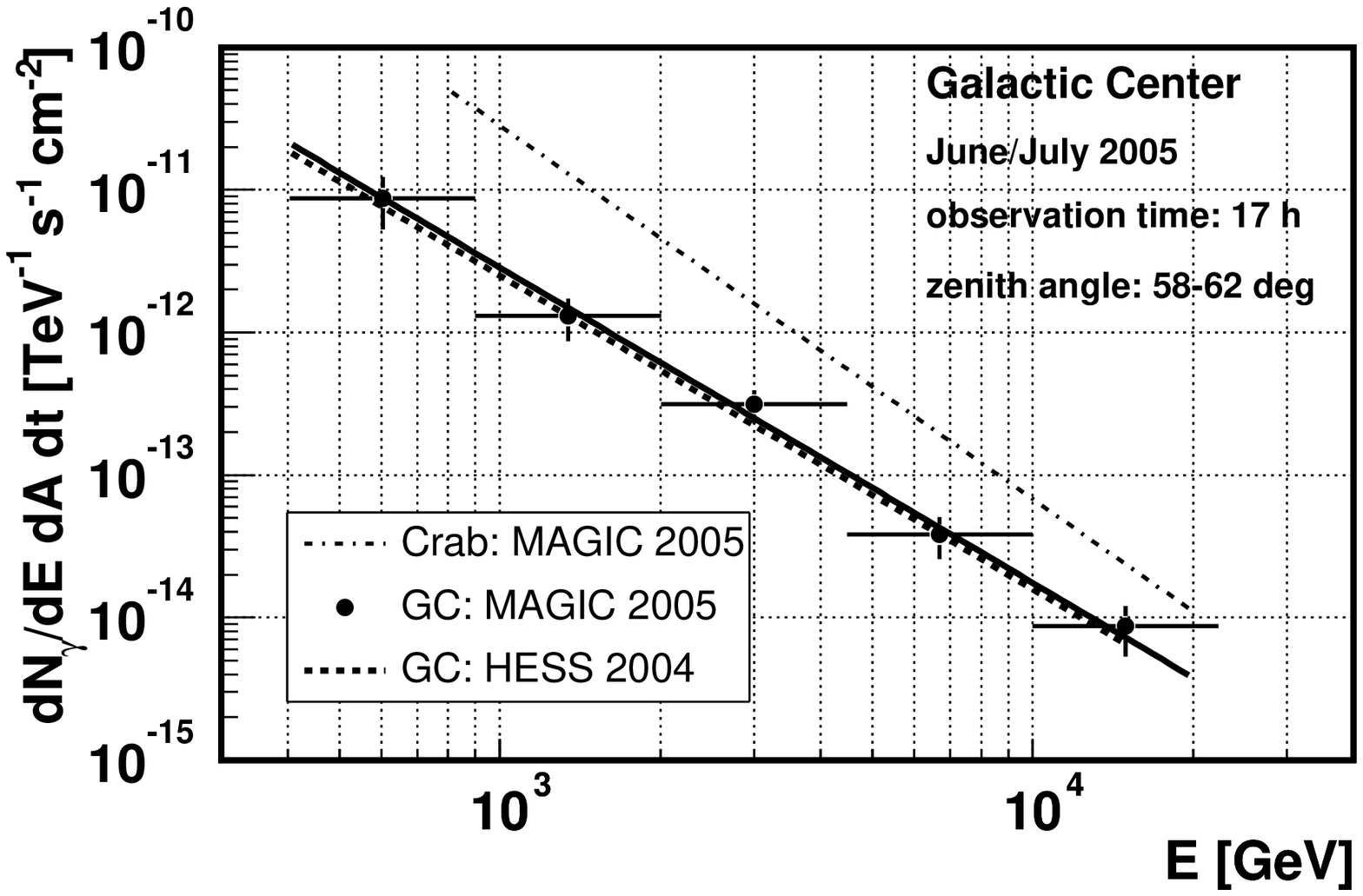,width=6cm}}
\end{center}
\caption{Left : Spectral energy density E$^2$dN/dE of gamma-rays
from HESS J1745-290 for 2003 (red empty circles) and 2004 (black
filled circles) datasets of the H.E.S.S. observation of the Galactic
Center. The shaded area shows the best power-law fit to the 2004
data points. The spectra expected from the annihilation of a
MSSM-like 14 TeV neutralino (dashed green line), a 5 TeV KK DM
particle (dotted pink line) and a 10 TeV DM particle annihilating
into 70\% $b\bar{b}$ and 30\% $\tau^+\tau^-$ in final state (solid
blue line) are presented. Right: Reconstructed VHE gamma-ray energy
spectrum of the GC (statistical errors only) as measured by the
MAGIC collaboration. The full line shows the result of a power-law
fit to the data points. The dashed line shows the 2004 result of the
HESS collaboration~\cite{Aharonian:2004wa}. The dot-dashed line
shows the energy spectrum of the Crab nebula as measured by MAGIC.}
\label{aba:gc}
\end{figure}

The MAGIC observations were carried out towards the Galactic Centre
since 2004 and revealed a strong emission. The observed excess in
the direction of the GC has a significance of 7.3 standard
deviations~\cite{Albert:2005kh} and is compatible with a pointlike
source. Large zenith observation angle ($\geq$60$^{\circ}$) implies
an energy threshold of $\sim$400 GeV. The source position and the
flux level are consistent with the measurement of
HESS~\cite{Aharonian:2004wa} within errors. The right hand side of
Fig.~\ref{aba:gc} shows the reconstructed VHE gamma-ray energy
spectrum of the GC after the unfolding with the instrumental energy
resolution. The differential flux can be well described by a power
law of index $\Gamma=2.2\pm0.2_{\rm stat.}\pm0.2_{\rm syst.}$. The
systematic error is estimated to be 35\% in the flux level
determination. The flux level is steady within errors in the
time-scales explored within these observations, as well as in the
two year time-span between the MAGIC and HESS observations.  An
interpretation in term of DM require a minimum value of mass higher
than 10 TeV. Most probably, if DM signal exists is overcome by other
astrophysical emitters.

\section{Dwarf Galaxies from the Local Group}\label{aba:dwarf}

Dwarf spheroidal galaxies in the Local Group are considered as
privileged targets for DM searches since they are among the most
extreme DM-dominated environments.  Measurements of roughly constant
radial velocity dispersion of stars imply large mass-to-luminosity
ratios. Nearby dwarfs are ideal astrophysical probes of the nature
of DM as they usually consist of a stellar population with no hot or
warm gas, no cosmic ray population and little dust. Indeed, these
systems are expected to have a low intrinsic gamma-ray emission.
This is in contrast with the Galactic Center where disentangling the
dominant astrophysical signal from possible more exotic one is very
challenging. Prior to the year 2000, the number of known satellites
was eleven. With the Sloan Digital Sky Survey (SDSS), a population
of ultra low-luminosity satellites has been unveiled, which roughly
doubled the number of known satellites. Among them are Coma
Berenices, Ursa Major II and Willman 1. IACTs have started
observation campaigns on dwarf galaxies for a few years.
Table~\ref{aba:tbl1} presents a possible list of the preferred
targets for DM searches.
\begin{table}
\tbl{A tentative list of preferred dwarf galaxies.}
{\begin{tabular}{@{}ccccc@{}}
\toprule
Name  & Distance$^{\text a}$ & Luminosity$^{\text a}$ &M/L$^{\text b}$ & Best positioned IACTs\\
&  (kpc)  & (10$^3$ L$_{\odot}$) & (M$_{\odot}$/L$_{\odot}$)   & \\
\colrule
Carina &101 &430 &40 &HESS, CANGAROO\\
Coma Berenices &44 &2.6 &450 &MAGIC, VERITAS\\
Draco &80 &260 &320 &MAGIC, VERITAS\\
Fornax& 138 &15500 &10 &HESS, CANGAROO\\
Sculptor &79 &2200 &7 &HESS, CANGAROO\\
Sagittarius &24& 58000$^{\text *}$& 25 &HESS, CANGAROO\\
Sextans &86 &500 &90 &HESS, CANGAROO\\
Ursa Minor &66 &290 &580 &MAGIC, VERITAS\\
Ursa Major II &32 &2.8 &1100& MAGIC, VERITAS\\
Willman 1 &38 &0.9 &700 &MAGIC, VERITAS\\
\botrule
\end{tabular}}
\begin{tabnote}
$^{\text a}$See Ref.~\refcite{Mateo:1998wg}.
$^{\text b}$See Ref.~\refcite{Simon:2007dq}.
$^{\text *}$See Ref.~\refcite{Helmi:2000qw}.\\
\end{tabnote}\label{aba:tbl1}
\end{table}

The star velocity dispersions in Draco reveal that this object is
dominated by DM on all spatial scales and provide robust bounds on
its DM profile, which thus decreases uncertainties on the
astrophysical contribution to the gamma-ray flux. The MAGIC
collaboration searched for a steady gamma-ray emission from the
direction of Draco~\cite{Albert:2007xg}. The analysis energy
threshold after cuts is 140 GeV. No significant excess is found. For
a power law with spectral index of 1.5, typical for a DM
annihilation spectrum, and assuming a pointlike source, the
2$\sigma$ upper limit is $\rm \Phi(E > 140 GeV) = 1.1 \times
10^{-11} cm^{-2} s^{-1}$. The measured flux upper limit is several
orders of magnitude larger than predicted for the smooth DM
distribution in mSUGRA models. The limit on the flux enhancement
caused by high clumpy substructures or a black hole is around
$\mathrm{O}(10^{3} - 10^{9})$. The WHIPPLE collaboration has also
observed Draco~\cite{Wood:2008hx}. Fig.~\ref{aba:globularcluster}
shows the 95\% C.L. exclusion curve on $\sigma v$. The upper limit
is at the level of 10$^{-22}$ cm$^{3}$s$^{-1}$ for 1 TeV neutralino
annihilating with BRs of 90\% in $b\bar{b}$ and 10\% in
$\tau^+\tau^-$.

The Sagittarius (Sgr) dwarf galaxy,  one of the nearest Galaxy
satellites of the Local Group, has been observed by H.E.S.S. since
2006. The annihilation signal from Sgr is expected to come from a
region of 1.5 pc, which is much smaller than the H.E.S.S. point
spread function. Thus, a pointlike signal has been searched for. No
significant gamma-ray excess is detected at the nominal target
position. A 95\% C.L. upper limit on the gamma-ray flux is derived:
$\rm \Phi^{95\%C.L.}_{\gamma} (E_{\gamma} > 250GeV) =
3.6\times10^{-12}cm^{-2}s^{-1}$, assuming a power-law spectrum of
spectral index 2.2~\cite{Aharonian:2007km}. A modelling of the Sgr
DM halo has been carried out. Two models of the mass distribution
for the DM halo have been studied: a cusped NFW profile and a cored
isothermal profile, to emcompass a large class of profiles. The
cored profile has a small core radius due to a cusp in the luminous
profile. The value of the line-of-sight-integrated squared density
is then found to be larger for the cored profile than for the NFW
profile (see Ref.~\refcite{Aharonian:2007km} for more details). The
left hand side of Fig.~\ref{aba:dwarf} presents the constraints on
the velocity-weighted annihilation cross section $\sigma v$ for a
cusped NFW and cored profiles in the solid angle integration
region $\rm \Delta\Omega = 2\times10^{-5}$ sr, for neutralino       
DM. Predictions for SUSY                
models are displayed. For a cusped NFW profile, H.E.S.S. does not
set severe constraints on $\sigma v$. For a cored profile, due to a
higher central density, stronger constraints are derived and some
pMSSM models can be excluded in the upper part of the scanned
region. Although the nature of Canis Major is stilll debated, HESS
observed this object and results on Canis
Major~\cite{Aharonian:2008dm} are presented in
Ref.~\refcite{mvivier}.
\begin{figure}   
\begin{center}
\mbox{\psfig{file=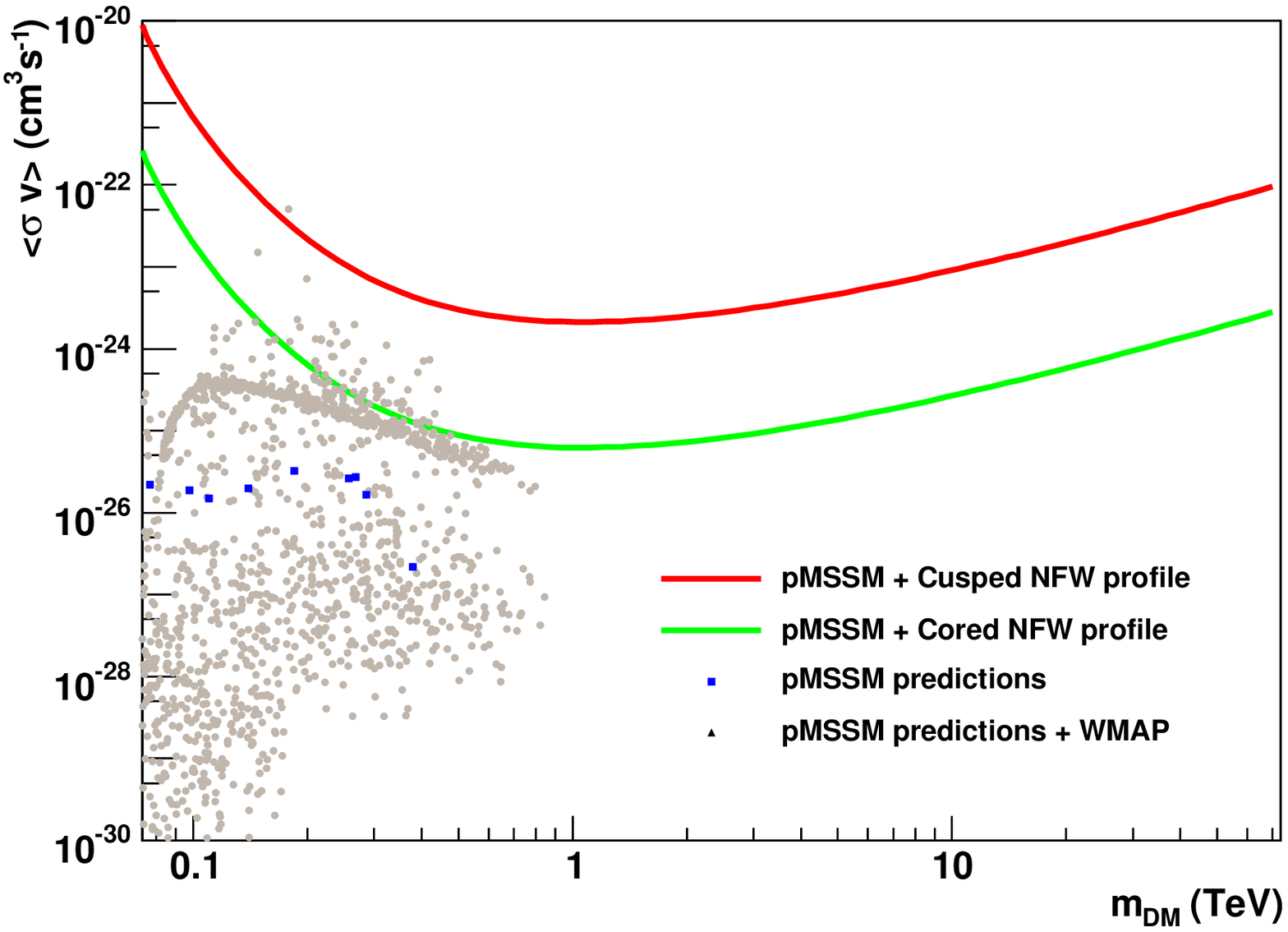,width=5.5cm,height=3.6cm}
\psfig{file=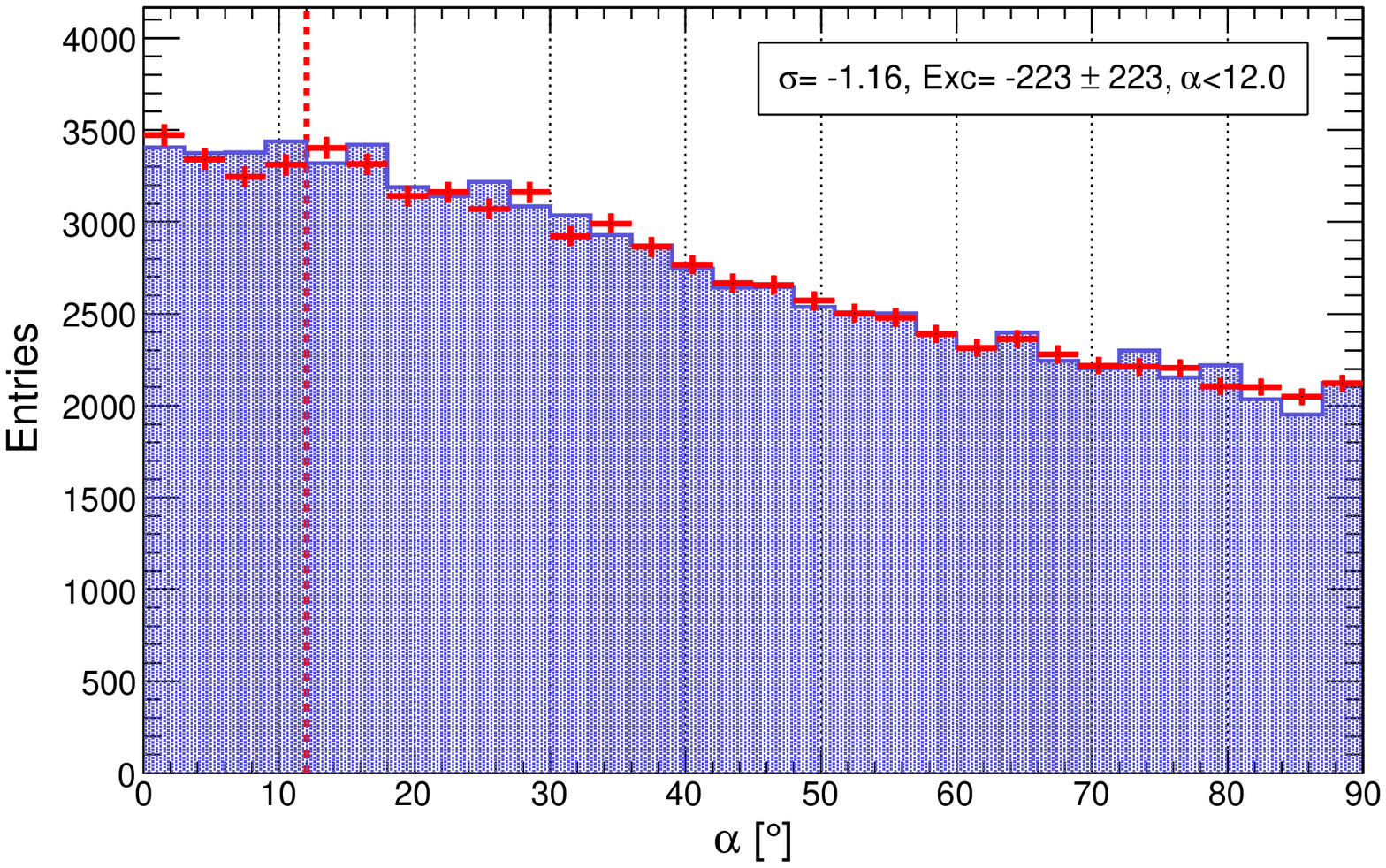,width=5.5cm}}
\end{center}
\caption{Left: Upper limits at 95\%C.L. on $\sigma v$ versus the
neutralino mass for a cusped NFW and cored DM halo profile for Sgr.
The predictions in pMSSM are also plotted with in addition those
satisfying in addition the WMAP constraints on the cold DM density.
  Right: Willman~1 $\alpha$-plot as seen by MAGIC in $15.5$~hours above
  a fiducial energy threshold of 100~GeV.
  The red crosses represent the ON--data sample, the blue
  shaded region is the OFF--data sample normalized to the ON--data
  sample between $30^\circ-80^\circ$. The vertical red dotted line
  represents the fiducial region $\alpha<12^\circ$ where the signal is
  expected.  }
\label{aba:dwarf}
\end{figure}

The recently discovered Willman 1 dwarf galaxy has been observed by
MAGIC~\cite{Aliu:2008ny}. Willman 1 has a a total mass of
$\sim$5$\times$10$^5$ M$_{\odot}$, which is  about an order of
magnitude smaller than those of the least massive satellite galaxies
previously known.  This object has one of the highest
mass-to-luminosity ratio. No significant gamma-ray excess beyond 100
GeV above the background was observed in 15.5 hours of observation
of the sky region around Willman 1. This is  shown in the right hand
side of Fig.~\ref{aba:dwarf}, where the
$^{\backprime\backprime}\alpha$-plot$^{\prime\prime}$ is
reported~\footnote{The $\alpha$-parameter is the angular distance
between the shower image main axis and the line connecting the image
barycenter and the camera center.}. The signal is searched with a
cut slightly larger than for a pointlike source to take into account
a possible source extension which makes a fiducial region to be
$\alpha$ $\leq$12$^{\circ}$. Flux upper limits of the order of
10$^{-12}$ cm$^{-2}$s$^{-1}$ for  benchmark mSUGRA models are
obtained~\cite{Aliu:2008ny}. Boost factors in flux of the order of
10$^3$ are required in the most optimistic scenarios, i.e. in the
funnel region or in region where the internal bremsstrahlung may
play an important role. However, uncertainties on the DM
distribution or the role of substructures may significantly reduce
this boost.

\section{Galactic Globular Clusters}\label{aba:globularclusters}

Globular clusters are not believed to be DM-dominated objects. Their
mass-to-luminosity ratios are well described by purely King profile
which suggets no significant amount of DM. However, the formation of
globular clusters fits in the hierarchical strucure formation
scenario in which globular clusters may have formed in DM
overdensities.  During their evolution, they may have hold some DM
in their central region. M15 is a nearby galactic globular cluster
with a core radius of 0.2 pc and a central density of
$\sim$10$^7$M$_{\odot}$pc$^{-3}$. Even if the mass-to-light ratio of
M15 does not suggest a significant component of DM, the compact and
dense core of M15 does not prevent from looking for a hypothetical
DM signal. The presence of stars and DM in globular clusters may
lead to a DM enhancement in their inner part. This is usually
treated with the adiabatic contraction
model~\cite{Blumenthal:1985qy}.

The Whipple collaboration obsered M15 in 2003. The modelling of DM profile
assuming adiabatic contraction of DM from an initial NFW distribution leads
to an enhancement in the
\begin{figure}   
\begin{center}
\psfig{file=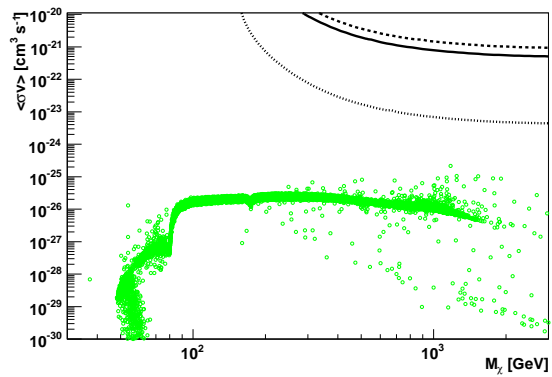,width=8cm}
\end{center}
\caption{Exclusion limit on $\sigma v$ as a function of the
neutralino mass m$_{\chi}$ on M15 (dotted line) assuming an
annihilation spectrum with 90\% in $b\bar{b}$ and 10\% in
$\tau^+\tau^-$, and the NFW profile after adiabatic contraction. The
solid line is the exclusion limit for Draco and the dashed line for
Ursa Minor.} \label{aba:globularcluster}
\end{figure}
astrophysical factor of $\sim$10$^2$ to 10$^3$.
Fig.~\ref{aba:globularcluster} shows the exclusion curve on $\sigma
v$ (dotted line) versus the neutralino mass assuming a NFW profile
after adiabatic contraction~\cite{Wood:2008hx}. The limit reaches
$\sim$10$^{-24}$ cm$^{3}$s$^{-1}$ for TeV neutralinos.

\section{Galaxy clusters}\label{aba:galaxyclusters}
Astrophysical systems such as the VIRGO galaxy
cluster~\cite{Baltz:1999ra} have been considered as target for DM
annihilation searches. The elliptical galaxy M87 at the center of
the VIRGO cluster has been observed by H.E.S.S. It is an active
galaxy located at 16.3 Mpc which hosts a black hole of
3.2$\times$10$^9$M$_{\odot}$. From 2003 to 2006, H.E.S.S. detected a
13$\sigma$ signal in 89 hour observation time at a location
compatible with the nominal position of the nucleus of
M87~\cite{Aharonian:2006ys}. With the angular resolution of
H.E.S.S., the VHE source is pointlike, with an upper limit on the
size of 3' at 99\% confidence level. Given the distance of M87, this
corresponds to a size of 13.7 kpc.
\begin{figure}     
\begin{center}
\psfig{file=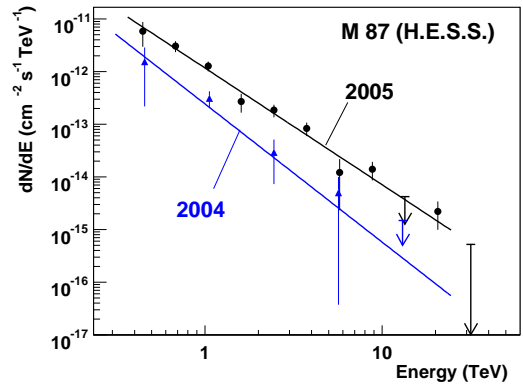,width=7cm}
\end{center}
\caption{Differential energy spectrum of M87 from $\sim$400
GeV up to 10 TeV from 2004 and 2005 data. Corresponding
fits to a power-law function dN/dE  E are plotted as
solid lines. The data exhibit indices of $\Gamma$ = 2.62$\pm$0.35 for
2004 and $\Gamma$ = 2.22$\pm$0.15 for 2005.}
\label{aba:m87}
\end{figure}

Fig.~\ref{aba:m87} shows the energy spectrum for 2004 and 2005 data.
Both data sets are well fitted to a power-law spectrum with spectral
indices $\Gamma$ = 2.62$\pm$0.35 (2004) and $\Gamma$ = 2.22$\pm$0.25
(2005). The 2005 average flux is found to be higher by a factor
$\sim$5 to the average flux of 2004. The integrated flux above 730
GeV shows a yearly variability at the level of
3.2$\sigma$~\cite{Aharonian:2006ys}. Variability at the day time
scale has been observed in the 2005 data at the level of 4$\sigma$.
This fast variability puts stringent constraints on the size of the
VHE emission region. The position centered on M87 nucleus excludes
the center of the VIRGO cluster and outer radio regions of M87 as
the gamma-ray emission region. The observed variability at the scale
of $\sim$2 days requires a compact emission region below 50R$_s$,
where R$_s\sim$10$^{-15}$ cm is the Schwarzschild radius of the M87
supermassive black hole~\cite{Aharonian:2006ys}. The short and long
term temporal variability observed with H.E.S.S. excludes the bulk
of the TeV gamma-ray signal to be of DM origin.

\section{Galactic Substructures: the Case for IMBHs}\label{aba:imbhs}

Mini-spikes around Intermediate Mass Black Holes have been recently
proposed as promising targets for indirect dark matter
detection~\cite{Bertone:2005xz}. The growth of massive black holes
inevitably affects the surrounding DM distribution. The profile of
the final DM overdensity, called mini-spike, depends on the initial
distribution of DM, but also on astrophysical processes such as
gravitational scattering of stars and mergers. Ignoring
astrophysical effects, and assuming adiabatic growth of the black
hole, if one starts from a NFW profile, a spike with a power-law
index 7/3 is obtained, as relevant for the astrophysical formation
scenario studied here characterized by black hole masses of
$\sim$10$^5$ M$_{\odot}$~\cite{Bertone:2005xz}. Mini-spikes might be
detected as bright pointlike sources by current IACTs.

\begin{figure}   
\begin{center}
\mbox{\hspace{-0.5cm}\psfig{file=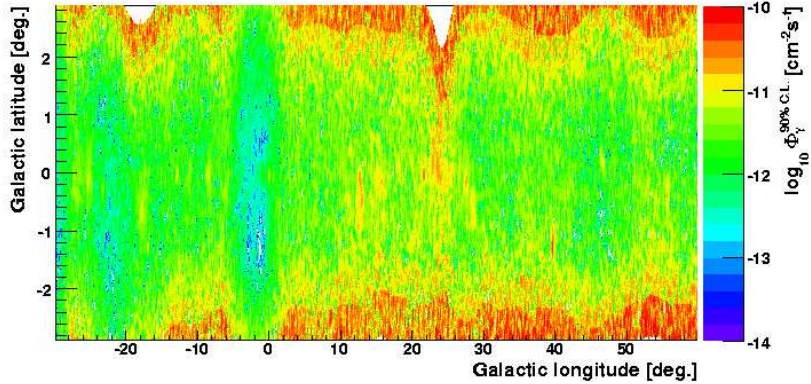,width=12cm}}
\end{center}
\caption{H.E.S.S. sensitivity map in Galactic coordinates, i.e. 90\%
C.L. limit on the integrated gamma-ray flux above 100 GeV, for dark
matter annihilation assuming a DM particle of mass m$_{\chi}$ = 500
GeV and annihilation into the $b\bar{b}$ channel. The flux
sensitivity is correlated to the exposure and acceptance maps. In
the Galactic latitude band between -2$^{\circ}$ and 2$^{\circ}$, the
gamma-ray flux sensitivity reaches 10$^{-12}$ cm$^{-2}$s$^{-1}$. }
\label{aba:sensitivity_hess}
\end{figure}

H.E.S.S. data collected between 2004 and 2007 during the Galactic
plane survey have allowed to accurately map the Galactic plane
between $\pm$3$^{\circ}$ in galactic latitude and from -30$^{\circ}$
to 60$^{\circ}$ in galactic longitude with respect to the Galactic Center position.
The study of the H.E.S.S. sensitivity in a large field of view to
dark matter annihilations has been
performed~\cite{Aharonian:2008wt}. Fig. ~\ref{aba:sensitivity_hess}
shows the experimentally observed sensitivity map in the Galactic
plane from Galactic longitudes l=-30$^{\circ}$ to l=+60$^{\circ}$
and Galactic latitudes b=-3$^{\circ}$ to b=+3$^{\circ}$, for a DM
particle of 500 GeV mass annihilating into the 100\% BR  $b\bar{b}$
channel.
The H.E.S.S. sensitivity depends strongly on the exposure time and
acceptance maps which are related to the choice of the pointing
positions. The flux sensitivity varies along the latitude and
longitude due to inhomogeneous coverage of the Galactic plane. In
the band between -2$^{\circ}$ and 2$^{\circ}$ in Galactic latitude,
a DM annihilation flux sensitivity at the level of 10$^{-12}$
cm$^{-2}$s$^{-1}$ is achieved for a 500 GeV DM particle annihilating
in the $b\bar{b}$ channel. Deeper observations of the Galactic
Center and at Galactic longitude of $\sim$20$^{\circ}$ allow the
flux sensitivity to be of $\sim$5$\times$10$^{-13}$
cm$^{-2}$s$^{-1}$. For b$\geq$2$^{\circ}$, the sensitivity is
deteriorated due to a weaker effective exposure. For b=0$^{\circ}$
and l=-0.5$^{\circ}$, the flux sensitivity is $\sim$10$^{-13}$
cm$^{-2}$s$^{-1}$ in $b\bar{b}$.

\begin{figure}  
\begin{center}
\mbox{\hspace{0cm}\psfig{file=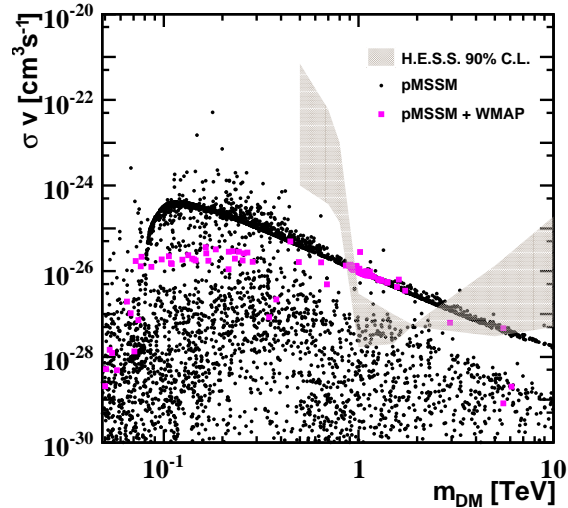,width=8cm}}
\end{center}
\caption{Constraints on the IMBH gamma-ray production scenario for
different neutralino parameters, shown as upper limits on $\sigma v$
as a function of the mass of the neutralino m$_{DM}$, but with a
number of implicit assumptions about the IMBH initial mass function
and halo profile (see Ref.~\refcite{Aharonian:2008wt} for details).
For the scenario studied here, the probability of having no
observable haloes in our Galaxy is 10\% from Poisson statistics
making these limits essentially 90\% C. L. exclusion limits for this
one particular (albeit optimistic) scenario (grey shaded area). The
DM particle is assumed to be a neutralino annihilating into
$b\bar{b}$ pairs or $\tau^+\tau^-$ pairs to encompass the softest
and hardest annihilation spectra. The limit is derived from the
H.E.S.S. flux sensitivity in the Galactic plane survey within the
mini-spike scenario. SUSY models (black points) are plotted together
with those satisfying the WMAP constraints on the DM particle relic
density (magenta points).} \label{aba:imbh_hess}
\end{figure}

H.E.S.S. reached the required sensitivity to be able to test DM
annihilations from mini-spikes in the context of one relatively
favorable scenario for IMBH formation and adiabatic growth of the DM
halo around the black hole (e.g. scenario B of
Ref.~\refcite{Bertone:2005xz}). H.E.S.S. observations (2004-2006) of
the Galactic plane allowed to discover more than 20 very high energy
gamma-ray sources~~\cite{Aharonian:2005kn}. Some of them have been
identified owing to their counterparts at other wavelengths, but
almost half of the sources have no obvious counterpart and are still
unidentified~\cite{Aharonian:2007zj}. An accurate reconstruction of
their energy spectra shows that all the spectra are consistent with
a pure power-law, spanning up to two orders of magnitude in energy
above the energy threshold. None of them exhibits an energy cutoff,
characteristic of DM annihilation spectra, in the energy range from
$\sim$100 GeV up to 10 TeV. Furthermore, the detailed study of their
morphology~\cite{Aharonian:2007zj} shows that all the sources have
an intrinsic spatial extension greater than $\sim$5 arcminutes,
while mini-spikes are expected to be pointlike sources for H.E.S.S.
No IMBH candidate has been detected so far by H.E.S.S. within the
survey range. Based on the absence of plausible IMBH candidates in
the H.E.S.S. data, constraints are derived on the scenario B of
Ref.~\refcite{Bertone:2005xz} for neutralino or LKP annihilations,
shown as upper limits on $\sigma v$~\cite{Aharonian:2008wt}.
Fig.~\ref{aba:imbh_hess} shows the exclusion limit at the 90\% C.L.
on $\sigma v$ as a function of the neutralino mass. The neutralino
is assumed to annihilate into $b\bar{b}$ and $\tau^+\tau^-$ with
100\% BR, respectively.
Predictions for SUSY models are also displayed. The limits on
$\sigma v$ are at the level of 10$^{-28}$ cm$^3$s$^{-1}$ for the
$b\bar{b}$ channel for neutralino masses in the TeV energy range.
Limits are obtained one mini-spike scenario and constrain on the
entire gamma-ray production scenario.

\section{Conclusion}

Dark matter searches will continue and searches with the phases 2 of
H.E.S.S. and MAGIC  will start soon. The phase 2  of H.E.S.S. will
consist of a new large 28 m diameter telescope located at the center
of the existing array, and will allow to lower the analysis energy
threshold down to less than 50 GeV. The installation of the second
telescope on the MAGIC site will allow to perform more sensitive
searches by the use of the stereoscopic mode. The upcoming
generation of IACTs, the Cherenkov Telescope Array, is in the design
phase. This array composed of several tens of telescopes will permit
to significantly improve the performances on both targeted DM
searches and wide-field-of-view DM searches.

\end{document}